\documentstyle[a4,11pt]{article}
\begin{document}
\begin{titlepage}
\vskip 2cm
\begin{flushright}
Preprint CNLP-1997-05
\end{flushright}
\vskip 2cm
\begin{center}
{\bf
SOLITON EQUATIONS IN  2+1 DIMENSIONS: REDUCTIONS, BILINEARIZATIONS
AND SIMPLEST SOLUTIONS}\footnote{Preprint
CNLP-1997-05.Alma-Ata.1997 \\
E-mail: cnlpmyra@satsun.sci.kz}
\end{center}
\vskip 2cm
\begin{center}
{\bf N.K.Bliev,  G.N.Nugmanova,
R.N.Syzdykova
and  R.Myrzakulov}
\end{center}
\vskip 1cm

\begin{center}
 Centre for Nonlinear Problems, PO Box 30, 480035, Alma-Ata-35, Kazakhstan
\end{center}

\begin{abstract}
In the preceding paper [13],
we presented, in particular, the  some (2+1)-dimensional integrable
reductions of the M-IX equation, moreover, their gauge equivalent
counterparts.
In this paper, we first construct the some new (1+1)-, and/or
(2+0)-dimensional  reductions - the $\sigma$- model equations
with potentials. We next establish the gauge equivalence between
the (1+1)-dimensional integrable classical compressible Heisenberg
ferromagnet models and the Yajima-Oikawa and Ma equations.
The bilinear forms of the M-IX and Zakharov equations and their
simplest one soliton solutions are found.
Also it is shown
that the Zakharov and Fokas equations are formally equivalent to each other.
\end{abstract}


\end{titlepage}

\setcounter{page}{1}
\newpage

\tableofcontents

\section{Introduction}

\quad  Considerable effort has been given recently to investigate (2+1) -
dimensional soliton equations (see, for example, [1-5]). The study of these
equations has thrown up new ideas in soliton theory and in other related
branches of mathematics and physics. Integrable spin systems are an
important both from the mathematical and physical points of view,
subclass of soliton equations [6-15]. In this paper we consider the
Myrzakulov IX (M-IX) equation [10]
$$
{\bf S}_t = {\bf S} \wedge M_1{\bf S} - iA_2{\bf S}_x - iA_1{\bf S}_y  \eqno(1a)
$$
$$
M_2u=2\alpha^{2} {\bf S}({\bf S}_x \wedge {\bf S}_y) \eqno(1b)
$$
where $ \alpha,b,a  $=  consts, ${\bf S} = (S_{1}, S_{2}, S_{3}),\quad
{\bf S}^{2} = E = \pm 1$ and
$$
M_1= \alpha ^2\frac{\partial ^2}{\partial y^2}+4\alpha (b-a)\frac{\partial^2}
   {\partial x \partial y}+4(a^2-2ab-b)\frac{\partial^2}{\partial x^2},
$$
$$
M_2=\alpha^2\frac{\partial^2}{\partial y^2} -2\alpha(2a+1)\frac{\partial^2}
   {\partial x \partial y}+4a(a+1)\frac{\partial^2}{\partial x^2},
$$
$$
A_1=i\{\alpha (2b+1)u_y - 2(2ab+a+b)u_{x}\},
$$
$$
A_2=i\{4\alpha^{-1}(2a^2b+a^2+2ab+b)u_x - 2(2ab+a+b)u_{y}\}.
$$

Equation (1) is  integrable in the sense that it admits the
Lax representation (LR) [10] and has  different types solutions [23].
In general
we will distinguish the two integrable cases: the M-IXA equation as
$\alpha^{2} = 1$ and the M-IXB equation as $\alpha^{2} = -1$. Besides,
equation (1) contains several interesting particular cases: (i)
$a = b = -1$,  yields the M-VIII equation;
(ii) $a = b = - \frac{1}{2}$, yields the celebrated Ishimori equation and
so on. Equation (1) is the (2+1)-dimensional
integrable generalisation  of the Landau-Lifshitz equation(LLE)
$$
{\bf S}_t = {\bf S} \wedge {\bf S}_{xx}  \eqno(2)
$$
and in 1+1 dimensions reduces to it.

This paper is a sequel to the preceding paper [13]. In [13],  we have
presented some results
on equation (1), in particular, the some (2+1)-dimensional reductions.
The main goal of the present
paper is to continue the
studies started in [13] and the key questions which we would like to
address here are the following ones. First of all we want to find
the other reductions in 1+1 and/or 2+0 dimensions.
Secondly, we will present the bilinear form of equation (1) that
allows us to construct the different exact solutions of (1) and we
demonstrate its work, presenting the simplest 1-soliton solution (1-SS).

The outline of this paper is as follows. After recalling in section 2,
some basic facts related to the M-IX equation,
in section 3 we  will present some (2+0)-dimensional reductions of
equation(1) - the some $\sigma$ -models with potentials and their
equivalent counterparts. In section 4 we show how derive  the
M-XXXIV equation from equation (1), which describe nonlinear dynamics of
compressible magnets. Also  we establish the
gauge equivalence between the M-XXXIV equation and the Yajima-Oikawa
equation (YOE). In sections 5 and 6, we present the bilinear forms of
equation (1) and its  gauge equivalent counterpart - the Zakharov
equation (ZE),  and use it to construct their simplest 1-SS. A connection
between the ZE  and the Fokas equation (FE) is disscused in section 7.
The last section is devoted to the concluding remarks.

\section{Some basic facts on the M-IX equation}

\quad In this section we briefly recall the some basic facts related to
the M-IX equation (1).

\subsection{The Lax representation}

As integrable, equation (1)  has the following LR [10]
$$ \alpha \Phi_y = [S+(2a+1)I]\Phi_x \eqno(3a) $$
$$ \Phi_t=2i[S+(2b+1)I]\Phi_{xx}+W\Phi_x \eqno(3b) $$
with
$$ W = 2i\{(2b+1)(F^{+} + F^{-} S) +(F^{+}S + F^{-}) +
(2b-a+\frac{1}{2})SS_x+\frac{1}{2}S_{x} + \frac{\alpha}{2} SS_y \}, \quad
$$
$$
S= \pmatrix{
S_3 & rS^- \cr
rS^+ & -S_3
},\quad S^{\pm}=S_{1}\pm iS_{2} \quad  S^2 = EI,\quad E = \pm 1,
\quad r^{2}=\pm 1,
$$
$$
F^{\pm} = A \pm D , \quad A=i[u_{y} - \frac{2a}{\alpha} u_{x}],
\quad D=i[\frac{2(a+1)}{\alpha} u_{x} - u_{y}].
$$
The compatibility condition of these linear equations gives
$$
iS_t + \frac{1}{2}[S, M_1 S] + A_2 S_x + A_1 S_y = 0  \eqno(4a)
$$
$$
M_2u = \frac{\alpha^{2}}{2i}tr( S[ S_x , S_y]) \eqno(4b)
$$
which is the matrix form of equation (1) within to change $t$ to $-t$.

\subsection{Gauge equivalent counterpart}

\quad It is well known that the gauge equivalent counterpart of equation (1)
is  the following ZE [5]
$$
iq_{t}+M_{1}q+vq=0 \eqno(5a)
$$
$$
ip_{t}-M_{1}p-vp=0  \eqno(5b)
$$
$$
M_{2}v = -2M_{1}(pq). \eqno(5c)
$$

This equation contains many interesting particular cases such as
the Davey-Stewartson (DS) equation, the YOE and so on. In 1+1
dimensions equation (5) reduces to the nonlinear
Schr${\ddot o}$dinger equation (NLSE)
$$
iq_{t}+ q_{xx}+2E\mid q\mid^{2}q=0. \eqno(6)
$$

The LR of the ZE (5) is given by [5]
$$
\alpha \Psi_y =2B_{1}\Psi_x + B_{0}\Psi \eqno(7a)
$$
$$
\Psi_t=4iC_{2}\Psi_{xx}+2C_{1}\Psi_x+C_{0}\Psi \eqno(7b)
$$
with
$$
B_{1}= \pmatrix{
a+1 & 0 \cr
0   & a
},\quad
B_{0}= \pmatrix{
0   &  q \cr
p   &  0
}
$$
$$
C_{2}= \pmatrix{
b+1 & 0 \cr
0   & b
},\quad
C_{1}= \pmatrix{
0   &  iq \cr
ip  &  0
},\quad
C_{0}= \pmatrix{
c_{11}  &  c_{12} \cr
c_{21}  &  c_{22}
}
$$
$$
c_{12}=i[2(2b-a+1)q_{x}+i\alpha q_{y}],\quad
c_{21}=i[2(a-2b)p_{x}-i\alpha p_{y}].
$$
Here $c_{jj}$ is the solution of the  following  equations
$$
2(a+1) c_{11x}- \alpha c_{11y} = i[2(2b-a+1)(pq)_{x} + \alpha (pq)_{y}]
\eqno (8a)
$$
$$
2ac_{22x}-\alpha c_{22y} = i[2(a-2b)(pq)_{x} - \alpha (pq)_{y}].
\eqno (8b)
$$

Note that the ZE (5) admits the Painleve property and is integrable
in this P-sense [15]. Also it has the different types solutions (solitons,
dromions and so on) [15, 27]

\subsection{The (2+1)-dimensional reductions}

\quad As above mentioned, equations (1) and (5) contain the several important
particular cases. Let us recall the some of these cases.

\subsubsection{The Myrzakulov VIII equation}

First, let us  we consider the reduction of the M-IX equation (4)
as $ a=b=-1$.
We have
$$
iS_t+\frac{1}{2}[S,S_{YY}]+iwS_Y = 0 \eqno(9a)
$$
$$
w_{X} + w_{Y} + \frac{1}{4i}tr(S[S_X,S_Y]) = 0 \eqno(9b)
$$
where $X=x/2, \quad Y = y/\alpha , \quad w = - \alpha^{-1}u_{Y}$.
This equation is the Myrzakulov VIII (M-VIII) equation [1].  The gauge
equivalent counterpart of equation (9) is given by
$$
iq_{t}+ q_{YY} + vq = 0 \eqno(10a)
$$
$$
ip_{t}-p_{YY} - vp = 0 \eqno(10b)
$$
$$
v_{X} + v_{Y} + 2(pq)_{Y} =0  \eqno(10c)
$$
which we denote as the M-VIII$_{q}$ equation.
The LR of these equations we can get from equations (3) and (7),
respectively, as $a=b=-1$ [10]. Equations (9)-(10) admit the different types
exact solutions, such as solitons, dromions, vortices  and so on [10, 24-25].

\subsubsection{The Ishimori equation }

Now let  $ a=b=-\frac{1}{2} $. Then equation (4) reduces to the known
Ishimori equation [9]
$$
iS_t+\frac{1}{2}[S,(S_{xx}+\alpha^{2}S_{yy})]+
iu_{y}S_x+iu_{x}S_y = 0 \eqno(11a)
$$
$$
\alpha^{2}u_{yy} - u_{xx}= \frac{\alpha^{2}}{2i}tr(S[S_x,S_y]). \eqno(11b)
$$

From equation (5) we get the  gauge equivalent counterpart of equation (11)
$$
iq_t + q_{xx} + \alpha^{2}q_{yy} + vq = 0 \eqno(12a)
$$
$$
\alpha^{2}v_{yy} - v_{xx}=-2(\alpha^{2}(pq)_{yy} + (pq)_{xx}) \eqno(12b)
$$
which is the famous  DS equation. This fact was for first time
established in [2].  The LR of (11) and (12)
we can get from
(3) and (7), respectively, as $ a = b = - \frac{1}{2}$.

\section{$\sigma$-models with potentials}

It is interesting to note that equation (1) admits some (1+1)-dimensional
(and/or (2+0)-dimensional) reductions. In this section we present the
$\sigma$-model with potential, which are the stationary limit of the
M-IX equation.

Consider   the (2+1)-dimensional LLE
$$
{\bf S}_{t} = {\bf S}\wedge \triangle {\bf S}
\eqno (13)
$$
It is well known  that the LLE (13) in the stationary limit coincide
with the  $\sigma$-model equation
$$
 \triangle {\bf S} + (\nabla {\bf S})^{2} {\bf S} = 0. \eqno (14)
$$
Here
$$
{\bf S}^{2} = S^{2}_{3} + r^{2} (S^{2}_{1} + S^{2}_{2}) = E = \pm 1, \quad
\triangle = \frac{\partial^{2}}{\partial x^{2}} +
\alpha^{2} \frac{\partial^{2}}{\partial y^{2}},
$$
$$
\nabla = {\bf i}\frac{\partial}{\partial x} +
{\bf j}\frac{\partial}{\partial y}, \quad {\bf i}^{2} = 1, \quad {\bf j}^{2}
 = \alpha^{2} = \pm 1, \quad
r^{2} = \pm 1, \quad E = \pm 1.
$$
From these results naturally arises the following question:
what $\sigma$-model equation is the stationary limit of equation (1)?.
Let us find it. In the stationary limit,
i.e. when  the spin vector ${\bf S}$ independent of $t$, equation (4)
takes the form
$$
M_1S + \{ a_{1} S^{2}_{x} + a_{2} S_{x}S_{y} + a_{3} S^{2}_{y} \} S
+ A_2 SS_x+A_1 SS_y = 0 \eqno(15a)
$$
$$
M_2u=\frac{\alpha^{2}}{2i}tr(S[S_x,S_y]) \eqno(15b)
$$
where $M_{1}$ we write in the form
$$
M_1= a_{3}\frac{\partial ^2}{\partial y^2}+a_{2}\frac{\partial^2}
   {\partial x \partial y}+a_{1}\frac{\partial^2}{\partial x^2},\quad
a_{1}=4(a^{2}-2ab-b), \quad a_{2}=4\alpha (b-a), \quad a_{3}=\alpha^{2},
$$
which is the Myrzakulov XXXII (M-XXXII) equation [10]. At the same time, the stationary limit
of the ZE (5) looks like
$$
M_{1}q+vq=0, \quad M_{1}p + vp=0, \quad M_{2}v = -2M_{1}(pq). \eqno(16)
$$
which  is the some complex modified  Klein-Gordon equation (mKGE).

Also we would like note that the M-XXXII equation (15),   in turn
contains the several particular cases. So, for example, in the case
$a=b=-\frac{1}{2}$,  we have the following $\sigma$-model
$$
S_{xx}+\alpha^{2}S_{yy} + (S^{2}_{x}+\alpha^{2}S^{2}_{y})S +
iu_{y}SS_x+iu_{x}SS_y = 0 \eqno(17a)
$$
$$
\alpha^{2}u_{yy} - u_{xx} = \frac{\alpha^{2}}{2i}tr(S[S_x,S_y]) \eqno(17b)
$$
which is the Myrzakulov XIII (M-XIII) equation [10].
In this case, from the mKGE (16) we obtain the  corresponding  equivalent
equation
$$
q_{xx}+\alpha^{2}q_{yy}+ vq = 0 \eqno(18a)
$$
$$
\alpha^{2}v_{yy} - v_{xx}=-2\{\alpha^{2}(pq)_{yy}
 +(pq)_{xx}\}. \eqno(18b)
$$

\section{Reductions in 1+1 dimensions: the gauge equivalence of
the Myrzakulov XXXIV equation and the YOE}

Now let us we consider the case, when $ X = t$. Then the M-VIII equation (9)
pass to the following Myrzakulov XXXIV (M-XXXIV) equation
$$
iS_t+\frac{1}{2}[S,S_{YY}]+iwS_Y = 0 \eqno(19a)
$$
$$
w_{t} + w_{Y} + \frac{1}{4}\{tr(S^{2}_{Y})\}_{Y} = 0. \eqno(19b)
$$

The  M-XXXIV  equation (19)
was proposed in [10] to describe nonlinear dynamics of
compressible magnets. It is integrable and has  the different soliton
solutions [10, 26].

In our case  equation (10)  becomes
$$
iq_{t}+ q_{YY} + vq = 0 \eqno(20a)
$$
$$
ip_{t}-p_{YY} - vp = 0  \eqno(20b)
$$
$$
v_{t} + v_{Y} + 2(pq)_{Y} =0  \eqno(20c)
$$
that is the YOE [16]. So, we have proved that the
M-XXXIV equation (19) and the YOE (20) is gauge equivalent
to each other.  The LR of (19) and (20) we can get from
(3) and (7), respectively, as $ a = b = - 1$ ( see, for example, the
ref.[10]). Note that our LR for the YOE (20) is different than that which was
presented in [16].

Also we would like note that the M-VIII equation (9) we usually write in
the following form
$$
iS_t + \frac{1}{2}[S, S_{\xi\xi}]+iwS_{\xi} = 0   \eqno (21a)
$$
$$
w_{\eta} - \frac{1}{4i}tr(S[S_{\xi},S_{\eta}]) = 0 \eqno (21b)
$$
where
$$
\xi = \frac{x}{2} + \frac{a+1}{\alpha}y, \quad \eta = -\frac{x}{2} -
\frac{a}{\alpha}y.    \eqno(22)
$$
This equation also admits the some soliton solutions [24].
The gauge equivalent counterpart of this  equation is given by
$$
iq_{t}+q_{\xi \xi}+vq=0 \eqno(23a)
$$
$$
v_{\eta} + 2r^{2}(\bar q q)_{\xi} = 0 \eqno(23b)
$$
that is the other ZE [5].
As $ \eta = t$  equation (21) take the other form of the M-XXXIV equation
$$
iS_t + \frac{1}{2}[S, S_{\xi\xi}]+iwS_{\xi} = 0   \eqno (24a)
$$
$$
w_{t} +  \frac{1}{4}(trS^{2}_{\xi})_{\xi} = 0.  \eqno (24b)
$$
Some soliton solutions of this equation were found in [26], using the
Hirota bilinear method. The gauge equivalent counterpart of the M-XXXIV
equation (24) we get from (23)
$$
iq_{t}+q_{\xi \xi}+vq=0, \eqno(25a)
$$
$$
v_{t} + 2r^{2}(\bar q q)_{\xi}=0, \eqno(25b)
$$
which is called the Ma equation (ME) and  was considered in [17]. The
LR of equations (24) and (25) were found in [10]. Note that our LR of
equation (25) is different than the LR, which was presented in [17].

\section{Bilinearization of the M-IX equation}

\quad The M-IX equaion contains rich class exact solutions such as solitons,
dromions, vortices, lumps and so on [10, 23]. To construct solutions of soliton
equations there are exist several powerful methods [1-5]. One of beatfull and
constructive among them is the Hirota method.  Here we make use of
Hirota's method to get exact 1-SS of equation (1), which
satisfies the boundary condition  ${\bf S} = (0, 0, 1)$
as $x,y = \pm \infty$. The use of the IST method is left in the future.
By putting [23]
$$
S^{+} = \frac{2\bar f g}{\bar f f + \bar g g}, \quad
S_3 = \frac{\bar f f - \bar g g}{\bar f f + \bar g g}, \eqno (26a)
$$
$$
u_x = 2i\alpha (2a+1)\frac{D_{x}(\bar f\circ f + \bar g \circ g)}{\bar f f +
\bar g g} - 2i\alpha^{2}\frac{D_{y}(\bar f \circ f +
\bar g \circ g)}{\bar f f + \bar g g}  \eqno (26b)
$$
$$
u_y = 8ia(a+1)\frac{D_{x}(\bar f \circ f + \bar g \circ g)}{\bar f f +
\bar g g}  - 2i\alpha (2a+1) \frac{D_{y}(\bar f \circ f +
\bar g \circ g)}{\bar f f + \bar g g}  \eqno (26c)
$$
the M-IX equation (1) is transformed into the bilinear equations [23]
$$
[iD_{t} - 4(a^{2} -2ab -b)D_{x}^{2} - 4 \alpha (b-a)D_{x} D_{y} -
\alpha^{2} D_{y}^{2} ](\bar f \circ g) = 0  \eqno (27a)
$$
$$
[iD_{t} - 4(a^{2} -2ab -b)D_{x}^{2} - 4 \alpha (b-a)D_{x} D_{y} -
\alpha^{2} D_{y}^{2} ](\bar f \circ f - \bar g \circ g) = 0  \eqno (27b)
$$
In addition, we have the following condition
$$
D_{y} \{2 \alpha (2a+1)D_{x} (\bar f \circ f + \bar g \circ g) -
2 \alpha^{2} D_{y} (\bar f \circ f + \bar g \circ g) \} \circ (\bar f f  +
\bar g g)=
$$
$$
D_{x} \{8a(a+1)D_{x} (\bar f \circ f + \bar g \circ g) -
2 \alpha(2a+1)D_{y} (\bar f \circ f + \bar g \circ g) \} \circ (\bar f f  +
\bar g g),
\eqno (28)
$$
which follows from the compatibility condition $ u_{xy} = u_{yx} $.

\subsection{Particular  cases}

\subsubsection{The M-VIII equation}
\quad In this case $a=b=-1$ and the expressions (26b,c)  take the form
$$
u_x = -2i\alpha \frac{D_{x}(\bar f \circ f + \bar g \circ g)}
{\bar f f + \bar g g}
- 2i\alpha^{2}\frac{D_{y}(\bar f \circ f + \bar g \circ g)}
{\bar f f + \bar g g}, \quad
u_y =  2i\alpha  \frac{D_{y}(\bar f\circ f + \bar g\circ g)}
{\bar f f + \bar g g}     \eqno (29a)
$$
or
$$
u_{X} = -2i\alpha \frac{D_{X}(\bar f \circ f + \bar g \circ g)}
{\bar f f + \bar g g}
- i\alpha\frac{D_{Y}(\bar f \circ f + \bar g \circ g)}
{\bar f f + \bar g g}, \quad
w =  - 2i  \frac{D_{Y}(\bar f\circ f + \bar g\circ g)}
{\bar f f + \bar g g}     \eqno (29b)
$$
where $w = -\alpha^{-1}u_{Y}$. Equation (27) becomes [24]
$$
[iD_{t}  - \alpha^{2} D_{y}^{2} ](\bar f \circ g) = 0 ,\quad
[iD_{t} - \alpha^{2} D_{y}^{2} ](\bar f\circ f - \bar g\circ g) = 0
\eqno (30a)
$$
or
$$
[iD_{t}  -  D_{Y}^{2} ](\bar f \circ g) = 0 ,\quad
[iD_{t} - D_{Y}^{2} ](\bar f\circ f - \bar g\circ g) = 0
\eqno (30b)
$$
In addition, we have from (28)
$$
D_{y} \{2 \alpha D_{x} (\bar f\circ f + \bar g \circ g) +
2 \alpha^{2} D_{y} (\bar f\circ  f + \bar g\circ g) \} \circ (\bar f f  + \bar g g)=
 - 2 \alpha D_{x}\{D_{y} (\bar f\circ f + \bar g\circ  g) \} \circ (\bar f f  + \bar g g)
\eqno (31a)
$$
or
$$
D_{Y} \{ D_{X} (\bar f\circ f + \bar g \circ g) +
2  D_{Y} (\bar f\circ  f + \bar g\circ g) \} \circ (\bar f f  + \bar g g)=
 -  D_{X}\{D_{Y} (\bar f\circ f + \bar g\circ  g) \} \circ (\bar f f  + \bar g g)
\eqno (31b)
$$
which are biquadratic.

\subsubsection{The Ishimori equation}

\quad  Let $a=b=-\frac{1}{2}$. Then the expressions (26) for the derivatives
of the potential $u$ become
$$
u_x = - 2i\alpha^{2}\frac{D_{y}(\bar f\circ f + \bar g \circ g)}{\bar f f +
\bar g g}, \quad
u_y = -2i\frac{D_{x}(\bar f\circ f + \bar g \circ g)}{\bar f f + \bar g g}.
\eqno (32)
$$
At the same time, equation (27) is transformed into the bilinear equations [9]
$$
[iD_{t} - D_{x}^{2} - \alpha^{2} D_{y}^{2} ](\bar f \circ g) = 0  \eqno (33a)
$$
$$
[iD_{t} - D_{x}^{2} - \alpha^{2} D_{y}^{2} ](\bar f \circ f -
\bar g \circ g) = 0.  \eqno (33b)
$$
In addition, we have the following biquadratic condition
$$
\alpha^{2} D_{y}\{D_{y} (\bar f \circ f + \bar g \circ g) \} \circ (\bar f f  + \bar g g)=
D_{x} \{D_{x} (\bar f\circ f + \bar g\circ g) \} \circ (\bar f f  + \bar g g)
\eqno (34)
$$
which follows from (28). Note that the equations (32)-(34) are the same
as in [9].

\subsubsection{The M-XXXIV equation}

\quad At last we consider the case $a=b=-1, x=2X =2t,
y=\alpha Y, w =-\alpha^{-1}u$. In this case we have
$$
u_t = - 2i\alpha \frac{D_{t}(\bar f\circ f + \bar g\circ g)}{\bar f f + \bar g g}
- i\alpha \frac{D_{Y}(\bar f\circ f + \bar g\circ  g)}{\bar f f + \bar g g}
\eqno (35a)
$$
$$
w=-\alpha^{-!}u_Y = -2i  \frac{D_{Y}(\bar f\circ f +
\bar g\circ g)}{\bar f f + \bar g g}.
\eqno (35b)
$$
The corresponding  bilinear equations are given by [26]
$$
[iD_{t} -  D_{Y}^{2} ](\bar f \circ g) = 0, \quad
[iD_{t} -  D_{Y}^{2} ](\bar f \circ f - \bar g \circ  g) = 0.  \eqno (36)
$$
The biquadratic condition has the form
$$
D_{Y} \{D_{t} (\bar f \circ f + \bar g\circ g) +  2
D_{Y} (\bar f\circ f + \bar g\circ g) \} \circ (\bar f f  + \bar g g)=
-D_{t}\{D_{Y} (\bar f\circ f + \bar g \circ g) \} \circ (\bar f f  + \bar g g).
\eqno (37)
$$

\subsection{Simplest soliton solution}

\quad In this section  we get, the simplest soliton solution
of equation (1) (for details, see, e.g. [23]). As example, we present  only
the 1-SS of the M-IXA equation, i.e. as $\alpha^{2}=1$.
The bilinear equation (27) represents the starting point
to obtain interesting classes of
solutions for the equation (1). The construction of the soliton
solutions is standard.
One expands the functions $g$ and $f$ as a series

$$g = \epsilon g_{1} + \epsilon^{3} g_{3} + \epsilon^{5}g_{5} +
\cdot \cdot \cdot \cdot \cdot, \eqno(38a)$$

$$f=1+\epsilon^2 f_2+\epsilon^4 f_4+\epsilon^6 f_6+ ..... \quad . \eqno(38b)$$

Substituting these expansions into (27) and equating the coefficients
of $\epsilon $, in the 1-SS case, one obtains the following system
of equations:

$$\epsilon^1: L(1\circ g_{1})=0 \eqno(39a)$$

$$\epsilon^3: L(\bar f_{2} \circ  g_1) = 0 \eqno(39b)$$
$$\epsilon^2: L(1\circ f_{2} + \bar f_{2} \circ 1 - \bar g_1 \circ g_1) = 0
\eqno(39c)$$

$$\epsilon^4: L(\bar f_{2} \circ f_{2}) = 0 \eqno(39d)$$
where
$$
L = iD_{t} - 4(a^{2} -2ab -b)D_{x}^{2} - 4 \alpha (b-a)D_{x} D_{y} -
\alpha^{2} D_{y}^{2} ](\bar f \circ g).   \eqno (40)
$$

We now ready  to construct the 1-SS of equation (1). In order
to construct exact 1-SS of equation (1), we take the ansatz

$$
g_1= \exp {\chi_1},\quad \chi_1 = m_1x + n_1y + c_1t + e_1 \eqno(41)
$$
where $m_{1}, n_{1}, c_{1}$ and $e_{1}$ are complex constants. By sustituting
the above value of $g_1$ in eq.(38a), we get

$$
c_1=i(a_{1}m_1^{2}+a_{2}m_{1}n_{1} + a_{3}n_{1}^{2}). \eqno(42)
$$

Substituting (41) in (39b,c,d), we obtain the expression for $f_2$ as

$$
f_2=B\exp{(\chi_1+\chi_1^*)} \eqno(43)
$$
where $ B = B_{R} + iB_{I}$ with
$$
B_{R} = \frac{1}{4}\{[2c_{1R}B_{I}-(a_{1}m_{I}^{2}+a_{2}m_{I}n_{I}+
a_{3}n_{I}^{2})][a_{1}m_{R}^{2}+a_{2}m_{R}n_{R} + a_{3}n_{R}^{2}]^{-1}
-1\},
$$
$$
B_{I} = \frac{1}{2}[\alpha^{2}n_{1R}n_{1I}-\alpha(2a+1)(m_{1I}n_{1R} +
m_{1R}n_{1I})
$$
$$
+4a(a+1)m_{1R}m_{1I}][2\alpha(2a+1)m_{1R}n_{1R}-
\alpha^{2}n^{2}_{1R}-4a(a+1)m_{1R}^{2}]^{-1}.
$$
The derivatives of potential have the forms
$$
u_{x} = \frac{K_{1}}{exp(-2\chi_{1R} +(2B_{R} +1)  +
 \mid B \mid^{2} exp(2\chi_{1R}} \eqno (44a)
$$
$$
u_{y} = \frac{K_{2}}{exp(-2\chi_{1R} +(2B_{R} +1)  +
 \mid B \mid^{2} exp(2\chi_{1R}} \eqno (44b)
$$
where
$$
K_{1} = 4\alpha (2a +1)  (2B_{I}m_{1R} +m_{1I}) -
4\alpha^{2}  (2B_{I}n_{1R} +n_{1I} $$
$$
K_{2} = 16a(a +1)  (2B_{I}m_{1R} +m_{1I}) -
4\alpha (2a+1)  (2B_{I}n_{1R} +n_{1I} $$

From the biquadratic condition (28) is obtained
$$
n_{1R} K_{1}=m_{1R} K_{2}.   \eqno(45)
$$

By substituting the above values of $g_1$ and $f_2$ in equations (26),
we obtain the expressions for the spin components and for the potential.
In detail, the different types of solutions (solitons, dromions,
lumps, vortices) of the M-IX equation were  presented  in  [10, 23].

\section{Bilinearization of the Zakharov  equation and its  1-SS}

To find the bilnear form of the ZE (5), we introduce the following
transformation
$$
q = \frac{G}{\phi}, \quad p = \frac{P}{\phi}. \eqno(46)
$$
Inserting this tranformation in equation (5), we get the Hirota
bilinear form of the ZE (5) as [27]
$$
[iD_{t} - 4(a^{2} -2ab -b)D_{x}^{2} - 4 \alpha (b-a)D_{x} D_{y} -
\alpha^{2} D_{y}^{2} ](G \circ \phi) = 0  \eqno (47a)
$$
$$
[iD_{t} - 4(a^{2} -2ab -b)D_{x}^{2} - 4 \alpha (b-a)D_{x} D_{y} -
\alpha^{2} D_{y}^{2} ](P\circ \phi) = 0  \eqno (47b)
$$
$$
[4a(a+1)D_{x}^{2} - 2\alpha (2a+1)D_{x} D_{y} + \alpha^{2} D_{y}^{2})
(\phi \circ \phi) = -2PG  \eqno (47c)
$$
with
$$
v=2M_{2}\log \phi .   \eqno(48)
$$

\subsection{Limiting cases}

\subsubsection{The M-VIII$_{q}$ equation (10)}

\quad In this case $a=b=-1$ and the bilinear equation (47) takes the form [25]
$$
[iD_{t}  - \alpha^{2} D_{y}^{2} ](G \circ \phi) =
[iD_{t}  -  D_{Y}^{2} ](G \circ \phi) = 0  \eqno (49a)
$$
$$
[iD_{t} - \alpha^{2} D_{y}^{2} ](P \circ \phi) =
[iD_{t} -  D_{Y}^{2} ](P \circ \phi) = 0  \eqno (49b)
$$
$$
(2\alpha D_{x} D_{y} + \alpha^{2} D_{y}^{2})(\phi \circ \phi)
= ( D_{X} D_{Y} + D_{Y}^{2})(\phi \circ \phi) = -2PG  \eqno (49c)
$$
with
$$
v = 2[\alpha^{2}\partial^{2}_{y} + 2\alpha \partial^{2}_{xy}]\log \phi
= 2[\partial^{2}_{Y} + \partial^{2}_{XY}]\log \phi .   \eqno(50)
$$

\subsubsection{The Davey-Stewartson equation}

\quad  Let $a=b=-\frac{1}{2}$. In this case  the bilinear equations become
$$
[iD_{t} - D_{x}^{2} - \alpha^{2} D_{y}^{2} ](G \circ \phi) = 0  \eqno (51a)
$$
$$
[iD_{t} - D_{x}^{2} - \alpha^{2} D_{y}^{2} ](P \circ \phi) = 0  \eqno (51b)
$$
$$
(-D_{x}^{2}  + \alpha^{2} D_{y}^{2})(\phi \circ \phi)
= -2PG  \eqno (51c)
$$
where $v=M_{1}^{\prime}\log \phi, \quad M_{1}^{\prime} = M_{1},$
as $a=b=-\frac{1}{2}$.

\subsubsection{The YOE}

\quad Now consider the case $a=b=-1, x=2X =2t$. In this case we have
$$
[iD_{t} -  D_{Y}^{2} ](G \circ \phi ) = 0  \eqno (52a)
$$
$$
[iD_{t} -  D_{Y}^{2} ](P\circ \phi) = 0  \eqno (52b)
$$
$$
( D_{t} D_{Y} +  D_{Y}^{2})(\phi \circ \phi)
= -2PG.  \eqno (52c)
$$
Note that in this case the potential is equal to
$$
v = 2[\partial^{2}_{Y} +  \partial^{2}_{tY}]\log \phi.    \eqno(53)
$$

\subsection{Simplest soliton solution of the ZE (5)}

\quad Let us, as example, we present  the 1-SS of the ZE (5)
 in the case $\alpha^{2}=1$ and $P=E\bar G$.
Note that equations (47) allow us to obtain the interesting classes of
solutions for the ZE (5) [27, 15]. The construction of the solutions is standard.
One expands the functions $G$ and $\phi$ as a series of $\epsilon$

$$
G = \epsilon G_{1} + \epsilon^{3} G_{3} + \epsilon^{5}G_{5} +
\cdot \cdot \cdot \cdot \cdot, \eqno(54a)
$$
$$
\phi =1+\epsilon^2 \phi_2+\epsilon^4 \phi_4+\epsilon^6 \phi_6 + .....\quad .
\eqno(54b)
$$

Substituting these expansions into (47) and equating the coefficients
of $\epsilon $, in the 1-soliton case, one obtains the following system
of equations:

$$
\epsilon^1: L(1\circ G_{1})=0  \eqno(55a)
$$
$$
\epsilon^3: L(\bar \phi_{2} \circ  G_1) = 0  \eqno(55b)
$$
$$
\epsilon^2: L(1\circ \phi_{2} + \phi_{2} \circ 1 +
2 \bar G_1 \circ G_{1}) = 0
\eqno(55c)
$$
$$
\epsilon^4: L(\phi_{2} \circ \phi_{2}) = 0  \eqno(55d)
$$
where
$L$ is given by (40). Using these equations  we can  construct the 1-SS
of equation (5). In order to construct exact 1-SS of equation (5), as above,
we take the ansatz
$$
G_1= \exp {\chi_1},\quad \chi_1 = m_1x + n_1y + c_1t + e_1 \eqno(56)
$$
where $m_{1}, n_{1}, c_{1}$ and $e_{1}$ are complex constants.
By sustituting
the above value of $G_1$ in equation (55a), we get

$$
c_1=i(a_{1}m_1^{2}+a_{2}m_{1}n_{1} + a_{3}n_{1}^{2}).
$$

From (56) and  (55b,c,d), we obtain the
expression for $\phi_2$ as
$$
\phi_2=B^{\prime}\exp{(\chi_1+\chi_1^*)},\eqno(57)
$$
where
$$
B^{\prime}  = -\frac{E}{3(b_{1}m_{1R}^{2}+b_{2}m_{1R}n_{1R}+b_{3}n_{1R}^{2}}.
$$

By substituting the above values of $G = G_1$ and $\phi = 1 + \phi_2$
in equations (46), we obtain the expressions for the field $q$ and
for the potential $v$.  This 1-SS and its generalizations and also
dromions, lumps, vortices types solutions  of the ZE (and the corresponding
solutions of the M-IX equation) we have considered, in detail,
in [15, 10, 27, 23]).

\section{Integrability: a connection between the ZE and the FE}

\quad In order to see whether equation (5) [and hence equation (1)],
is in general integrable, in [15] we have  carried out the  singularity
structure analysis of equation (5) and shown that it [and (1)] has the
Painleve property. Here this statement we prove using the following
observation: the ZE and the FE are formally equivalent to each other.
For this
purpose we write equation (4) in terms of the coordinates $\xi, \eta$ as
$$
iS_t +  \frac{1}{2}[S,(b+1) S_{\xi \xi} -bS_{\eta \eta}] +
ibw_{\eta} S_{\eta} + i(b+1)w_{\xi}S_{\xi} = 0 \eqno(58a)
$$
$$
w_{\xi \eta} =  \frac{1}{4i}tr(S[S_{\xi},S_{\eta}]). \eqno(58b)
$$
This equation, for convenience, in [10] we called the M-XX equation.
Its   gauge equivalent equation looks like
$$
iq_t+ (1 + b)q_{\xi \xi } - b q_{\eta \eta } + vq = 0 \eqno(59a)
$$
$$
ip_t - (1 + b)p_{\xi \xi } + b q_{\eta \eta } -  vq = 0 \eqno(59b)
$$
$$
v_{\xi \eta } = -2\{(1+ b) (pq)_{\xi \xi} - b(pq)_{\eta \eta}\}. \eqno(59c)
$$
Equations (58) - (59) are of course integrable in the sense that admit
the LR and have the different types solutions (solitons, dromions,
lumps and so on) [15, 23, 27]. In fact these
equations are not new, and are only the new forms of equations (4) and (5),
respectively in terms of $\xi,  \eta$, which are given by (22).

Now we make the simplest scaling tranformation: from $(t,\xi, \eta, q,p,v)$
to $(Ft, C\xi, D\eta, $\\
$Aq, Bp, F^{-1}v)$. Then, for example,  equation (59)
takes the form
$$
iq_t - (\gamma - \beta)q_{\xi \xi } +  (\gamma + \beta)
q_{\eta \eta } + vq = 0 \eqno(60a)
$$
$$
ip_t + (\gamma - \beta)p_{\xi \xi } -  (\gamma + \beta)
p_{\eta \eta } - vq = 0 \eqno(60b)
$$
$$
v_{\xi \eta } = -2\lambda
[(\gamma + \beta)(pq)_{\eta\eta } -  (\gamma - \beta)
(pq)_{\xi\xi}] \eqno(60c)
$$
where
$$
\lambda =ABCD,
\quad C^{2} = \frac{(\gamma +\beta)(b+1)}{(\gamma - \beta)b},
\quad F = \frac{(\beta - \gamma)}{b+1}C^{2}
$$
$$
\gamma = -\frac{1}{2}F[(b+1)D^{2}+bC^{2}]C^{-2}D^{-2}, \quad
\beta = \frac{1}{2}F[(b+1)D^{2}-bC^{2}]C^{-2}D^{-2}.
$$

Now let us we consider the FE [4]
$$
iq_t - (\gamma - \beta)q_{\xi \xi } +  (\gamma + \beta)
q_{\eta \eta } - 2\lambda q[(\gamma +
\beta)(\int^{\xi}_{-\infty}(pq)_{\eta}d\xi^{\prime}
$$
$$
+ v_{1}(\eta, t))
-  (\gamma - \beta)(\int^{\eta}_{-\infty}(pq)_{\xi} d\eta^{\prime}
+ v_{2}(\xi,t))] =0 \eqno(61a)
$$
$$
ip_t + (\gamma - \beta)p_{\xi \xi } -  (\gamma + \beta)
p_{\eta \eta } + 2\lambda p[(\gamma +
\beta)(\int^{\xi}_{-\infty}(pq)_{\eta}d\xi^{\prime}
$$
$$
+ v_{1}(\eta, t))
-  (\gamma - \beta)(\int^{\eta}_{-\infty}(pq)_{\xi} d\eta^{\prime}
+ v_{2}(\xi,t))] =0 \eqno(61b)
$$
with $ p=\bar q$ and  in contrasr with the equation (60), in our case
$\xi, \eta$ are the characteristic coordinates defined by
$$
\xi = x+y, \quad \eta = x-y. \eqno(62)
$$

This equation also contains several  interesting particular cases. Let
us recall these cases.

(i) $\gamma = \beta = \frac{1}{2}, v_{1} = v_{2} =0,$ yields equation
$$
iq_t +q_{xx} - 2\lambda q\int^{y}_{-\infty}(pq)_{x}dy^{\prime}
= 0, \quad \lambda = \pm 1.  \eqno(63)
$$
As noted by Fokas, equation (63) is perhaps the simplest complex
scalar equation in 2+1 dimensions, which can be solved by the IST method.
It is also worth pointing out that when $x=y$ this equation reduces
to the NLSE (6).

(ii) $\gamma = 0, \beta = 1$, yields the celebrated DSI equation
$$
iq_t + q_{\xi \xi } +  q_{\eta \eta }
- 2\lambda q[(\int^{\xi}_{-\infty}(pq)_{\eta}d\xi^{\prime} +
v_{1}(\eta, t))
+(\int^{\eta}_{-\infty}(pq)_{\xi} d\eta^{\prime}
+ v_{2}(\xi,t))] =0 .\eqno(64)
$$
This equation has the Painleve property  and admits exponentially
localized solutions including dromions for nonvanishing boundaries.

(iii) $\gamma = 1, \beta =0$ yields the DSIII equation
$$
iq_t - q_{\xi \xi } +  q_{\eta \eta }
- 2\lambda q[(\int^{\xi}_{-\infty}(pq)_{\eta}d\xi^{\prime} +
v_{1}(\eta, t))
-(\int^{\eta}_{-\infty}(pq)_{\xi} d\eta^{\prime}
+ v_{2}(\xi,t))] =0. \eqno(65)
$$
Equation (65)  also supports certain localized solutions.

Now let us return to equation (61) and  introduce the potential $V$ by
$$
V = - 2\lambda [(\gamma +
\beta)(\int^{\xi}_{-\infty}(pq)_{\eta}d\xi^{\prime} + v_{1}(\eta, t))
-  (\gamma - \beta)(\int^{\eta}_{-\infty}(pq)_{\xi} d\eta^{\prime}
+ v_{2}(\xi,t))]. \eqno(66)
$$
Then the FE (61) takes the form
$$
iq_t - (\gamma - \beta)q_{\xi \xi } +  (\gamma + \beta)
q_{\eta \eta } + V q = 0  \eqno (67a)
$$
$$
ip_t + (\gamma - \beta)p_{\xi \xi } -  (\gamma + \beta)
p_{\eta \eta } + Vp = 0  \eqno(67b)
$$
$$
V_{\xi \eta } = -2\lambda
[(\gamma + \beta)(pq)_{\eta\eta } -  (\gamma - \beta)
(pq)_{\xi\xi}]. \eqno(67c)
$$

Comparing the ZE in the form (60) and the FE in the form (67), we see
that they have formally the same forms. Recently it was proved by
Radha and Lakshmanan [3] that the FE (61) satisfies the Painleve property and
hence it is expected to be integrable. From these results follow that
the ZE and hence its equivalent counterpart the M-IX equation
also satisfy the Painleve property and are integrable and in
this sense (also, see [15]). Of course, strictly speaking, this statement
is correct in the case when $\xi, \eta$ are real, i.e. when  $\alpha $
is real.
In particular, this is why the ZE contains and at the same time
the FE not contains the DSII equation.

\section{Conclusion}

The some new (1+1)-, and/or (2+0)-dimensional integrable reductions
of the M-IX equation and their equivalent counterparts are considered.
In particular,
we have established the gauge equivalence between the (1+1)-dimensional
integrable inhomogeneous continuous Heisenberg ferromagnets (the M-XXXIV
equation) and the YOE and ME.

We have also constructed the bilinear forms of the M-IX equation and the ZE
and of their reductions. Moreover the simplest 1-SS of the ZE and the
M-IX equation are found. Of course, these equations admit the generalizations
of these solutions and other interesting solutions such as dromions, vortices,
lumps and so on [15, 23, 27].

Also  we have shown that   the ZE and the FE are formally equivalent
to each other. As shown by Radha and Lakshmanan [3], the FE satisfies the
Painleve property, i.e. it is expected to be integrable. Hence and from
the results of [15] follow that
the ZE and its equivalent the M-IX equation are integrable in the Painleve
property sense.

Concluding, we note that between the some above considered equations as well
as between the other spin systems and the NLSE-type equations take place
the so-called  Lakshmanan equivalence or   L-equivalence. This problem we will
consider elsewhere (see, for example, the refs.[18-22]).

\section{Acknowledgments}

RM wishes  to thank Prof. M.Lakshmanan for
hospitality during his visits
to  Bharathidasan University and for useful discussions.
RM is grateful for helpful conversations with  J.Zagrodsinsky,
R. Choudhury,  A.Kundu, Radha Balakrishnan,  M.Daniel and R.Radha.  Also he  would like
to thank Dr. Radha Balakrishnan for her hospitality during  visits
to the Institute of Mathematical Sciences.

\end{document}